\documentclass[12pt,preprint]{aastex}

\shorttitle{Spectral shape of black hole binaries}
\shortauthors{Kubota et al. }
\begin{document}

\title{On the temperature profile of radiatively efficient geometrically thin disks in 
black hole binaries with the {\it ASCA} GIS}

\author{Aya Kubota\altaffilmark{1}, Ken Ebisawa\altaffilmark{2}, 
Kazuo Makishima\altaffilmark{1,3}, Kazuhiro Nakazawa\altaffilmark{4}}

\altaffiltext{1}{Institute of Physical and Chemical Research (RIKEN), 
2-1 Hirosawa, Wako, Saitama 351-0198, Japan }
\altaffiltext{2}{code 662, Laboratory of High Energy Astrophysics, NASA/Goddard Space Flight 
Center, Greenbelt,  MD 20771, U.S.A}
\altaffiltext{3}{Also Department of Physics,  University of Tokyo, 
7-3-1 Hongo, Bunkyo-ku,\\ Tokyo 113-0033, Japan}
\altaffiltext{4}{Institute of Space and Astronautical Science, 3-1-1 Yoshinodai,
 Sagamihara, Kanagawa 229-8510, Japan}
\email{aya@crab.riken.jp}

\begin{abstract}
X-ray spectra of black hole binaries in the standard high/soft state were
studied comprehensively by using {\it ASCA} GIS data, and partially {\it RXTE} PCA data. 
A mathematical disk model 
was applied to several
black hole binaries to see if the observed accretion disk
temperature profile was consistent with that expected from 
the standard accretion disk model.
This model is called $p$-free disk, and assumes 
that the spectrum is composed of 
multi-temperature blackbody emission of which 
the local temperature $T(r)$ at radius $r$ is given by
$T(r)\propto r^{-p}$ with $p$ being a positive free parameter. 
The standard disk,
where gravitational energy of the accreted matter is effectively released as blackbody
radiation,  
roughly requires $p \approx 3/4$,
while a small deviation is expected depending on the inner boundary conditions, 
general relativistic effects and disk vertical structures.
Our sample objects included
LMC X-1, LMC X-3, XTE J$2012+381$, and  GRO J$1655-40$.
During the {\it ASCA} observations, these black hole binaries showed characteristics of 
the standard high/soft state.  
Under the standard modeling of high-state black hole binaries, 
the sources show disk inner temperatures of 0.76--1.17~keV, 
the disk fraction to the total 0.7--10~keV flux 
of 54--98\%, and absorption columns of (0.7--12)$\times10^{21}~{\rm cm^{-2}}$. 
The best-fit values of $p$ were found in the range of 0.6--0.8,
and the standard value of $p=3/4$ was accepted for all the sources.
The obtained values of $p$ are also compared with those expected for the standard accretion disk 
in the Schwarzschild metric by using the so-called {\sc grad} model.
The observed $p$-values were indeed found to be consistent with 
those expected from the standard accretion disk in the Schwarzschid metric.

\end{abstract}


\thispagestyle{headings}

\section{Introduction}
From the classical perspective,
spectral states of black hole binaries are classified into two classes 
(e.g., Tanaka~1997 and references therein). 
One is the low/hard state in which X-ray spectra are represented by a 
single power-law with an exponential cut-off around several tens of keV to 100~keV.
This low state is realized when the mass accretion rate is relatively low.
The other is the high/soft state, which is known as a high accretion-rate state. 
In the high/soft state, X-ray spectra consist of a dominant soft component and a 
power-law tail. The dominant soft component is believed to be from 
the radiatively efficient, optically thick, geometrically thin disk  
(the "standard disk", Shakura \& Sunyaev 1973)
reaching the innermost stable orbit around a central black hole.
In fact, when the soft component is fitted with a "multi-color disk" model
(MCD model or {\sc diskbb} in {\sc xspec}\footnote{http://heasarc.gsfc.nasa.gov/docs/xanadu/}; 
Mitsuda et al.~1984) that approximates the
standard accretion disk by ignoring its inner boundary condition, 
the obtained inner radius is kept remarkably constant against any large 
intensity variation (e.g., Makishima et al.~1986; Tanaka~1989; Takizawa~1991; 
Ebisawa et al.~1993, 1994). 
Moreover, in many black hole binaries, including Cyg~X-1 (Dotani et al.~1997), 
once
the true inner radius, $R_{\rm in}$, is estimated by the MCD fit, 
it is found to be consistent with 
$R_{\rm ms}=6R_{\rm g}(=6GM/c^2$, with black hole mass $M$),
which is the last stable orbit for a non-spinning black hole, 
Here, from an apparent inner radius, $r_{\rm in}$, one of two fit parameters of the MCD, 
the true inner radius, $R_{\rm in}$, can be calculated as  
$R_{\rm in}=r_{\rm in}\cdot \kappa^2\cdot \xi$ 
with an inner boundary correction factor, $\xi=0.412$ (Kubota et al.~1998), 
and a correction factor of spectral hardening, $\kappa=1.7$--2.0 (Shimura \& Takahara 1995).

After this standard view was established, 
recent spectral studies of black hole binaries have revealed that there seems to be
three sub-statuses in the high/soft states: the standard regime, anomalous regime 
(or "weak very high state"; see below), 
and apparently standard regime, in the order of increasing the mass accretion rate 
(Kubota, Makishima \& Ebisawa 2001; Kubota \& Makishima~2004). 
The standard regime corresponds to the classical high/soft state where the inner radius is kept constant
under canonical modeling of the MCD plus power-law.
The anomalous regime is characterized by thermal inverse Compton scattering by high-temperature 
electrons around the optically thick disk that extends down to $R_{\rm ms}$,  
and by the existence of type-B QPOs (e.g., van der Klis~2004; Remillard et al.~2002). 
This anomalous regime is basically the same as a very high state
(VHS; Miyamoto et al.~1991), which was redefined as a steep power-law state
by McClintock \& Remillard (2003), and thus the anomalous regime was 
renamed as a weak VHS by Kubota \& Done (2004). 
Though the soft disk component no longer dominates the 
hard spectral component, 
the standard disk picture is still justified for the underlying optically thick disk
in this anomalous regime (Kubota et al.~2001; Kubota \& Makishima~2004). 
Together with these three stable sub-states in the high state, 
there seems to be an intermediate case, 
called an intermediate state in the literature, or a strong VHS by Kubota \& Done (2004),
at an occasion of the low-to-high transition.
%

The spectral shape of the apparently standard regime, the most luminous regime, 
resembles that of the 
classical high/soft state, and consists of the 
dominant soft component and a very weak power-law tail. 
However, the apparent disk inner radius, $r_{\rm in}$, obtained under the MCD fit 
is {\it not} constant, but varies as  $\propto T_{\rm in}^{-1}$, where $T_{\rm in}$ is the 
highest color temperature of the disk, while it is kept constant in the standard regime.
Plainly, the data in the apparently standard regime are {\em not} consistent with a constant inner
radius, constant emissivity and constant colour temperature correction, which are 
justified in the standard regime.
We here define this inconstancy of  $r_{\rm in}$ in the apparently standard regime 
as a deviation from the simple standard picture. 
This deviation is now commonly seen in black hole binaries in the most luminous case, 
including  GRO J$1655-40$ (Kubota, Makishima, \& Ebisawa~2001) and 4U~$1630-47$ (Abe et al.~2005).
%
Beyond our Galaxy and Large Magellanic Cloud (LMC), the 
apparently standard regime  
has also been suggested in off-nuclear 
ultra-luminous ($3\times 10^{39}\sim10^{40}~{\rm erg~s^{-1}}$) 
compact X-ray sources in nearby galaxies (hereafter ULXs; Makishima et al.~2000; Fabbiano~1989). 
That is to say, the ULXs often show the MCD-type spectra, and the values of $r_{\rm in}$ are not
kept constant, but proportional to $T_{\rm in}^{-1}$ (Mizuno, Kubota \& Makishima 2001), 
like black hole binaries in the apparently standard regime.

\begin{figure}[htbp]
\begin{center}
\includegraphics[angle=-90,scale=.40]{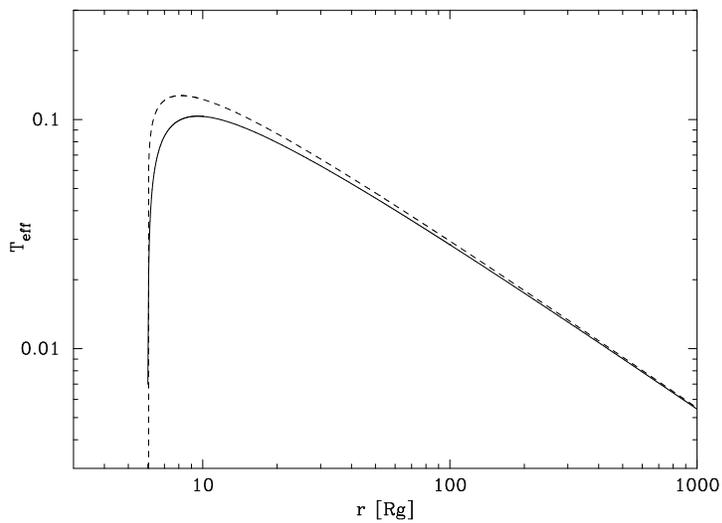}
\caption{Radial profiles of $T_{\rm eff}$ shown based on the standard accretion disk under the Newtonian potential (dashed line) and the Schwarzschild potential (solid line).}
\label{fig:tr}
\end{center}
\end{figure}

\begin{figure}[htbp]
\begin{center}
\includegraphics[angle=-90,scale=.4]{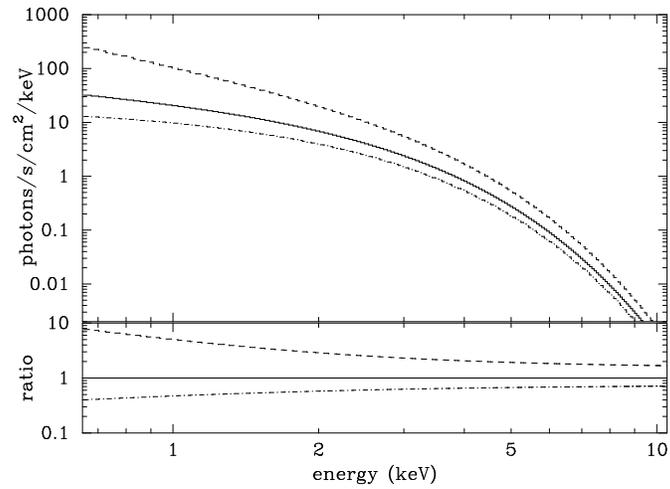}
\caption{Calculated spectra of the $p$-free disk model. 
The cases of $p=0.75$ (solid line), 0.5 (dashed line) and 
1.0 (dash-dot line) are shown under the assumption of 
$kT_{\rm in}=0.8$ keV and 
$(r_{\rm in}/{\rm km})^2\cos i/(d/10{\rm kpc})^2=10^4$.
The ratios to each models to the disk model of $p=0.75$ are also shown. }
\label{fig:mo}
\end{center}
\end{figure}

This deviation can be interpreted as a break of the standard disk (Kubota \& Makishima 2004), 
or can also be caused by a different color temperature correction (Gierli$\acute{\rm n}$ski \& Done 2004). 
As is done for XTE~J$1550-564$ by Kubota \& Makishima (2004), 
the break of the standard disk 
is considered in detail by investigating the local temperature gradient of the disk with 
a so-called $p$-free disk model (a local model for {\sc xspec}; Kubota \& Makishima 2004), 
with which the spectrum consists of the sum of blackbody emission with a local temperature of 
$T(r)=T_{\rm in}\cdot (r/r_{\rm in})^{-p}$ ($r$ is a distance from a central object).
In the Newtonian metric, the local effective temperature of the standard disk is given as
\begin{equation}
T_{\rm eff}(r)=\left\{\frac{3GM\dot{M}}{8\pi\sigma r^3}\cdot\left(1-\sqrt{\frac{R_{\rm in}}{r}}\right)\right\}^{1/4}~~~,
\label{eq:tr_n}
\end{equation}
with the Stefan-Boltzmann constant, $\sigma$,  
and the mass accretion rate, $\dot{M}$ (Shakura \& Sunyaev 1973).
Figure~\ref{fig:tr} exemplifies the temperature profile of the standard disk. 
Thus, the $p$-value is expected to be $3/4$ by ignoring the boundary condition at 
$r\approx R_{\rm in}$. 
This is the case of the MCD model (Mitsuda et al. 1984).
With another cooling process, the $p$-value is predicted to become smaller than $3/4$  
(e.g., Watarai et al.~2000).
Figure~\ref{fig:mo} exemplifies the 0.7--10~keV spectra of the $p$-free disk model 
with $p$=0.5, 0.75(=3/4) and 1.0. 
This figure shows that the $p$-free disk with a smaller value of $p$ 
reveals an X-ray spectrum in which the soft energy band is more 
enhanced than the standard disk of $p=3/4$. 

However, as already pointed out by Kubota \& Makishima~(2004), 
even in the standard accretion disk, the observed value of $p$ is expected to be 
slightly different from $3/4$ due to
(1) the disk inner boundary condition (see Fig.~\ref{fig:tr}), (2) general relativistic effects,
and (3) the disk vertical structure including electron scattering.
Therefore, to discuss the break of the standard disk, 
knowledge of the temperature profile of the standard disk must be required.
It is thus important to obtain an observationally acceptable range of $p$ 
in well-known black hole binaries in the standard high/soft state (standard regime), where the 
radiatively efficient standard disk is justified.
%
With this perspective, this paper presents a detailed study of the spectral shape of the standard-state
Galactic/Magellanic black hole binaries observed with the 
Gas Imaging Spectrometer (GIS; Ohashi et al.~1996; Makishima et al.~1996) 
onboard {\it ASCA}. The {\it ASCA} GIS 
has a fine spectral capability in the range of 0.7--10~keV, and 
is free from pile-up, even for the bright (X-ray flux of $\sim1$~Crab) Galactic objects.
Therefore, the GIS is 
the best detector 
to examine the structure of the optically thick 
accretion disk around bright Galactic/Magellanic black hole binaries in detail.
In this paper, we discuss the spectral shape 
by using the $p$-free disk model, following previous reports.

Before the present work, 
information of our sample is summarized in \S~2.
In \S~3 and \S~4, the observed data are analyzed, and the results are summarized. 
In \S~5, the observational result 
is briefly discussed in comparison with the standard disk picture and the latest models 
for the radiatively efficient, geometrically thin disks.


\placefigure{fig:tr}

\placefigure{fig:mo}

\section{Our sample and observations}

Many black hole binaries were observed with {\it ASCA} 
in their high/soft state  (e.g., Cyg X-1, 
GRS $1009-45$,  
LMC X-1, LMC X-3,  
GRO J$1655-40$, 
GRS $1915+105$, 
XTE J$2012+381$, 
XTE~J$1550-564$, 
and 4U~$1630-47$). 
In this paper,  the {\it ASCA} GIS spectra of 
LMC X-1, LMC X-3, XTE J$2012+381$ and GRO J$1655-40$ are reanalyzed based on 
the $p$-free disk model. 
They all show characteristics of the standard high/soft state, and 
show an inner temperature of $T_{\rm in}>0.7$ keV, 
ensuring that  
their Wien shoulders of local diluted 
blackbody radiation at the inner edge of the  disk  
appear at $>2$ keV. 
In addition, the absorption columns of $N_{\rm H}$ are not too high 
($N_{\rm H}\le10^{22}~{\rm cm^{-2}}$) to study the disk spectra in detail.
The {\it ASCA} data of XTE~J$1550-564$ are not included, because they 
characterize the VHS (Kubota \& Done 2004) rather than the classical high/soft state. 
In Table~\ref{tab:obs}, the observation logs of these four sources are shown with 
the estimated central object mass, inclination angles, $i$, and distances, $d$.
The large mass of the central objects in LMC~X-3, LMC~X-1, and GRO~J$1655-40$ 
fiducially measured from binary motion,
makes these three objects confirmed black holes (e.g., McClintock and Remillard 2003), 
and X-ray data of XTE~J$2012+381$ 
characterize its nature as an accreting black hole in the classical high/soft state.

For each source, the GIS events were extracted from a circular region of $6'$ radius 
centered on the image peak, after selecting good time intervals in 
a standard procedure. Average energy spectra were constructed from the two GIS detectors.
For the two Galactic sources, XTE~J$2012+381$ and GRO~J$1655-40$, 
the high bit-rate data were analyzed after correcting the dead time in reference to the
count-rate  monitor data (Makishima et al.~1996). 
The medium bit-rate data were not used, because
even the monitor scaler was overflown and a dead-time correction was not possible.
A systematic error of 1\% is included to each energy bin of the spectra, 
in order to account for calibration uncertainties.  
Because the two LMC sources are much dimmer, 
both the high bit-rate data and the medium bit-rate data are utilized, 
and the systematic error is not included.
A background was completely negligible compared to the fluxes of the Galactic sources, 
and thus was not subtracted.
For LMC sources, the background is not negligible, and is constructed 
from a source-free region of the GIS field of view.

There were simultaneous {\it RXTE} observations corresponding to the {\it ASCA} observations of  
XTE~J$2012+381$ and GRO~J$1655-40$, on 1998 May 29 and 1997 February 26 respectively.
The {\it RXTE} PCA data are very useful to estimate the power-law contribution, which 
is critical to examine the spectral shape of the dominant soft component with 
the $p$-free disk model.
We used {\sc heasoft}~5.2 in which systematic difference in calibration of GIS and PCA pointed
e.g., by Done, Madejski, Zycki~(2000) was corrected.
Thus, for these two sources, the  {\it RXTE} PCA data 
were simultaneously analyzed with the {\it ASCA} GIS data. 
For the PCA data reduction, a top layer was used from all five units, 
using standard exclusion criteria (target elevation less than $10^\circ$ above the limb of the Earth; 
pointing direction more than 1.2 arcmin from the target; data acquired within 30 minutes after 
spacecraft passage through South Atlantic Anomaly). 
The background was estimated for each observation using the software package {\sc pcabackest} (version 2.1e), supplied by the RXTE Guest Observer's Facility at NASA/GSFC. 
A response matrix was made for each observation by utilizing 
the software package {\sc pcarsp} version 8.0. 
In order to take into account the calibration uncertainties, a systematic error of 1\% was added 
to each bin of the PCA spectra. 
The 3--20 keV data were used in the present work, 
because there is sometimes a residual structure 
in the 20--35 keV range associated with the Xe-K edge at 30 keV.
 
\section{Analyses and Results}
\subsection{XTE J$2012+381$}



Responding to the discovery by the All Sky Monitor (Levine et al.~1996) 
on board {\it RXTE},
{\it ASCA} observed XTE J2012+381
on 1998 May 29--30 with a net exposure of 38 ks (White et al. 1998).
As previously shown by White et al.~(1998), the energy spectrum was very soft, 
and was successfully described with the canonical MCD plus power-law model.
Together with the high flux and low fractional contribution from 
the power-law tail, this fact makes the object ideal for the present study. 
The {\it RXTE} pointing observations of this source were performed 24 times; 
Vasiliev, Trudolyubov \& Revnivtsev~(2000) reported that the inner disk radius
is kept constant under the 
canonical MCD plus power-law model.
One of the {\it RXTE} observations on 1998 May 29 17:48:00--18:46:56 (UT) was simultaneous with 
the {\it ASCA} observation.
The GIS spectrum was reconstructed by using the data during this window with an exposure of 2.4~ks. 
The 3--20 keV {\it RXTE} PCA spectrum and the 0.7--10 keV {\it ASCA} GIS spectrum were jointly 
analyzed. 

\begin{figure}[htbp]
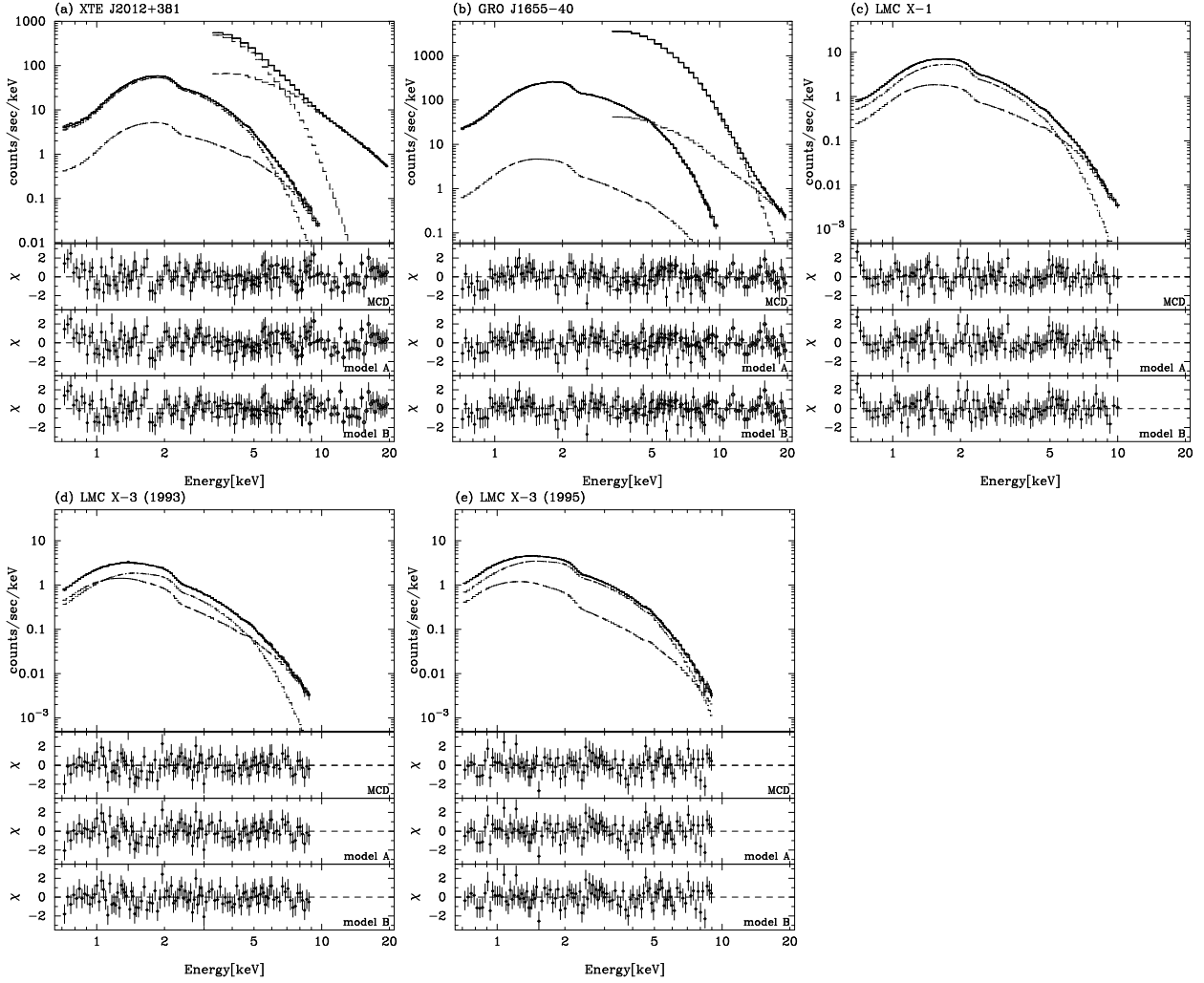

\includegraphics[angle=0,scale=.31]{f3a.eps}
\includegraphics[angle=0,scale=.31]{f3b.eps}
\includegraphics[angle=0,scale=.31]{f3c.eps}
\includegraphics[angle=0,scale=.31]{f3d.eps}
\includegraphics[angle=0,scale=.31]{f3e.eps}
\caption{GIS and PCA spectra of XTE J$2012+381$ ($a$), GRO J$1655-40$ ($b$), 
LMC X-1 ($c$), LMC X-3 in 1993 ($d$) and LMC X-3 in 1995 ($e$). 
The best-fit canonical MCD plus power-law model are superposed on the data. 
The residuals between the data and the canonical model, model $A$ and model $B$ are shown 
in the bottom three panels.}
\label{fig:spec}
\end{figure}

The overall spectrum was first examined with the MCD plus power-law model. 
Except for a normalization factor, all of the model parameters were constrained to be the same
between the GIS and the PCA data.
Both these components were assumed to undergo 
absorption with a common column density, $N_{\rm H}$.
The model reproduced the simultaneous  data very well with
$\chi^2/\nu=152.1/152$ on the condition that 
a Fe-K edge around 7.6~keV was included in the power-law component. 
The Fe-K edge structure was significant 
in terms of the $F$-test ($F(3,152)=218$), 
and the smeared-edge model ({\sc smedge} in {\sc xspec}; 
Ebisawa et al. 1994) gave an acceptable fit to the structure. 
Figure~\ref{fig:spec}$a$ 
shows the raw data and the residuals between the data and the model.
In Table~\ref{tab:result_j2012}, the best-fit parameters are shown together with 
the fraction of the disk luminosity to the total luminosity in the range of 0.7--10~keV. 
Though the values of $d$ and $i$ are not known, 
the true inner radius, $R_{\rm in}$, and the disk bolometric luminosity, $L_{\rm disk}$, 
were calculated with the assumptions of $d=10$~kpc and $i=60^\circ$.
Here and hereafter, a spectral hardening factor of $\kappa=1.7$ and a boundary correction factor 
of $\xi=0.412$ are used to estimate of $R_{\rm in}$.

The successful fitting of the canonical model 
means that the temperature gradient does not conflict with $p=3/4$, 
which is the case of the MCD.  However, 
to obtain both the best-fit value and an acceptable range of $p$, 
the soft component was next represented with the $p$-free disk model 
instead of the MCD model (hereafter model $A$). 
Other fitting conditions were the same as in the canonical fitting. 
As also shown in Table~\ref{tab:result_j2012}, 
the best-fit $p$ was determined as $0.71\pm0.02$ within 90\% confidence. 
Even though the power-law parameters have been 
accurately constrained thanks 
to the {\it RXTE} data, our basic assumption that the hard component has a 
single power-law shape down to $\sim1$ keV would not be warranted; 
the value of $p$ might change if these systematic uncertainties were taken into account.
This uncertainty was considered by 
applying an additional absorption colmun, $N_{\rm H}^{\rm hard}$, to the  
power-law component alone (hereafter model $B$). 
The results are also summerized in Table~\ref{tab:result_j2012}. 
The value of $p$ is obtained as $0.8\pm0.1$ with large
$N_{\rm H}^{\rm hard}$ of $2.5\pm0.8 \times10^{24}~{\rm cm^{-2}}$.
The bottom two panels in Fig.~\ref{fig:spec}$a$ show the residuals between the data and models 
$A$ and $B$. 

\begin{figure}[htbp]
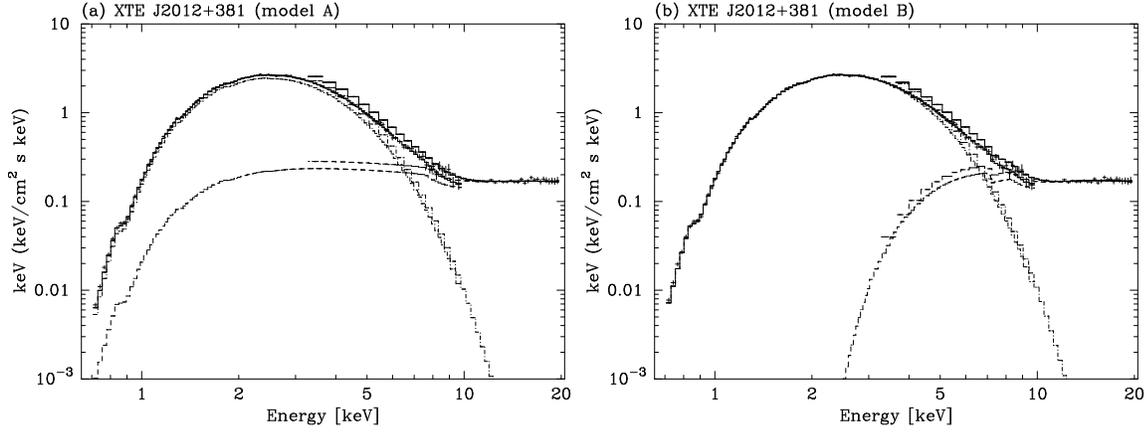

\includegraphics[angle=-90,scale=.31]{f4a.eps}
\includegraphics[angle=-90,scale=.31]{f4b.eps}
\caption{Unfolded spectra of XTE~J$2012+381$ with the best-fitting 
model~$A$ (panel $a$) and model~$B$ (panel $b$).}
\label{fig:spec_eeu}
\end{figure}

Figure~\ref{fig:spec_eeu} shows unfolded spectra with the best-fit predictions of 
models~$A$ and $B$.  
This figure shows that the 
power-law contribution becomes maximum with model~$A$, while it becomes minimum
with model~$B$, and thus the fittings with these two models give
the most conservative range for $p$.
The acceptable range of $p$ is thus determined to be
0.7--0.9, even when taking into acount the systematic uncertainties of the hard component. 
Hereafter, all of the sources are considered based on with the canonical model, 
model $A$ and model $B$.
%

\subsection{GRO J$1655-40$}
The X-ray transient GRO J$1655-40$ (Nova Sco 1994) is a secure black hole binary 
system with dynamical mass evidence of $M=6.0$--6.6$~M_\odot$ 
that exhibits superluminal jets  (e.g., Hjellming \& Rupen 1995; van der Hooft et al. 1998; Shahbaz
et al. 1999; Hannikainen et al.~2000). 
Based on multiple {\it RXTE} pointings of the 1996--1997 outburst of this source, 
Kubota et al.~(2001) showed that 
it exhibited the previously described three spectral regimes of a 
high luminosity state
with the critical disk luminosity of 1--2$\times10^{38}~{\rm erg~s^{-1}}$.

{\it ASCA} observations of GRO~J$1655-40$ were performed five times from 1994 to 1997.
The source was found in the high state in 
1995 August 15--16 and 1997 February 25--28 
(Zhang et al.~1997; Ueda et al.~1998; Yamaoka et al.~2001).
Referring to the spectral parameters of the 1996--1997 outburst (Kubota et al.~2001), 
on one hand, the source was found on the highest end of the standard regime during
the 1997 observation. 
On the other hand, the spectral parameters of the 1995 observation (e.g., Zhang et al.~1997)
reveal the characteristics of the anomalous regime (weak VHS), and thus 
the strong inverse Comton scattering may change the spectral shape from direct disk emission. 
Thus, we focus our discussion on the 1997 observation.
The {\it ASCA} observation of 1997 February 26 was simultaneously performed with 
{\it RXTE} (e.g., Yamaoka et al.~2001).
Net exposures of 2.4 ks of the {\it ASCA} GIS high-bit rate data 
and 5.4 ks of the {\it RXTE} PCA data were obtained after the standard data reductions.

Following the case of XTE~J$2012+381$, 
the 0.7--10 keV GIS spectrum and 
the 3--20 keV PCA spectrum were simultaneously analyzed with the canonical MCD plus 
power-law model, model $A$,  and model $B$.
All of the models assume that 
the soft disk component (MCD or $p$-free disk) and the 
power-law components are commonly modified by 
the inter-stellar absorption and other absorption features.
The fit results are shown in Table~\ref{tab:result_j1655} and 
Fig.~\ref{fig:spec}$b$.
In this spectral fitting, a negative Gaussian and three-edge structures 
were included. 
By referring to Yamaoka et al.~(2001), 
the center energy of the negative Gaussian was fixed at 6.81~keV, 
and the energies of the first two edges were fixed at 7.71 keV and 8.81 keV.  
The obtained values of the equivalent width of the negative Gaussian, 
and the optical depth of the edges are consistent with those shown in 
Yamaoka et al.~(2001).
In our joint fit, the other edge structure was required at $\sim10.4$~keV of 
great significance ($F(2,150)=156.9$ in the case of the canonical model).
Due to the high temperature of the disk, and very low intensity of the 
power-law component  (see Fig.~\ref{fig:spec}$b$),
the value of $\Gamma$ of the power-law could not 
be precisely determined even with the PCA data. 
Therefore, under a fitting with models $A$ and $B$, 
the value of $\Gamma$ was limited to the range of 2.0--2.5, which 
is the typical value for the power-law component in the standard high/soft state.
 

With the canonical fit, the true inner radius was obtained as 
$R_{\rm in}=22.6^{+0.2}_{-0.3}$~km, 
for $d=3.2$~kpc and $i=70^\circ$. 
The disk luminosity was obtained as 
$L_{\rm disk}=9\times10^{37}~{\rm erg~s^{-1}}$, 
which is roughly 10\% of the Eddington luminosity of a 6.0--6.6~$M_\odot$ black hole.
The confidence ranges of $R_{\rm in}$ and $L_{\rm disk}$ given in Table~\ref{tab:result_j1655} 
include systematic uncertainties of the
estimation of $d$ together with statistic errors. 
With models $A$ and $B$, 
the acceptable range of $p$ was obtained as 0.68--0.73.

\subsection{LMC X-1}
One of the first identified X-ray sources in the 
Large Magellanic Cloud, LMC X-1  
(Johnston, Bradt, \& Doxsey 1979), 
is known as a secure black hole binary with $M$=4--10~$M_\odot$ and $i<63^\circ$. 
This object was observed with {\it ASCA} on 1995 April 2--3 with a net exposure of 
21.7 ks. Makishima et al. (2000) reported that the {\it ASCA} GIS spectrum was well reproduced by 
the canonical MCD plus power-law model. 
The same data were reanalyzed based on the same condition for the reduction in 
Makishima et al. (2000). 
The fitting results with the canonical model, model $A$, and model $B$ 
are given in Table~\ref{tab:result_lmcx1} and in Fig.~\ref{fig:spec}$c$. 
Same as the case of XTE~J$2012+381$, the {\sc smedge} is significant with 
$F(3, 110)=31.1$.
With $d=50$ kpc 
(Freedman et al.~2001)
the true inner radius was obtained as 
$R_{\rm in}=59\pm2 \cdot (\cos i/\cos 30^\circ)^{-1/2}$ km from the canonical model. 
For a 4--10~$M_\odot$ black hole, this is consistent with $\sim6R_{\rm g}$.
The calculated value of $L_{\rm disk}$, 
$\sim 1.5\times10^{38}\cdot (\cos i/\cos 30^\circ)^{-1}~{\rm erg~s^{-1}}$, 
is 9--20\% of the Eddington limit for a 4--10 ~$M_\odot$ black hole with
an inclination of $i<63^\circ$.
The acceptable range of $p$ was determined to be 0.7--1.1 with models $A$ and $B$.

\subsection{LMC X-3}
Together with LMC~X-1, LMC X-3 is a persistent black hole binary in LMC with 
$M$=5.9--9.2~$M_\odot$.
It was observed with {\it ASCA} twice on 1993 September 22--23 and 
1995 April 14--15, with net exposures of 20.6 ks and 18.4 ks, respectively. 
The {\it ASCA} results have already been reported by Makishima et al.~(2000).
The same data as in Makishima et al.~(2000) were reanalyzed by applying the $p$-free disk model 
in this paper. 
The best-fit parameters based on the canonical model are given in Table~\ref{tab:result_lmcx3}. 
The raw data are shown with the best-fit canonical model in Fig.~\ref{fig:spec}$d$-$e$.
Without any edge structure, the canonical model has given 
acceptable fits to 0.7--9 keV GIS spectra, with 
$\chi ^2/\nu =102.2/109$ and $103.8/109$, for 1993 and 1995 observations, 
respectively, and the fitting parameters
are consistent with those from Makishima et al. (2000).
With $d=50$~kpc and $i=66^\circ$, 
the innermost radii and the disk bolometric luminosities were calculated 
to be $R_{\rm in}=47\pm2$ km and $46\pm2$ km, and 
$L_{\rm disk}=8.5\times10^{37}~{\rm erg~s^{-1}}$ 
and $1.7\times10^{38}~{\rm erg~s^{-1}}$ for each observation. 
The obtained $R_{\rm in}$ are 
consistent with $6R_{\rm g}$, 
and the $L_{\rm disk}$ values are roughly 6--10\% and 12--19\% of 
the Eddington luminosity of 5.9--9.2~$M_\odot$ BH. 
The $p$-free disk model was also applied to this source as model $A$ and model $B$,
and the acceptable ranges of $p$ were 
determined to be 0.51--0.75 and 0.60--0.70, for the 1993 and 1995 observations, respectively.


\placefigure{fig:spec}
\placefigure{fig:spec_eeu}

\section{Discussion}

\subsection{Summary of the observational results}
 
\begin{figure}[htbp]
\begin{center}
\includegraphics[scale=.4, angle=-90]{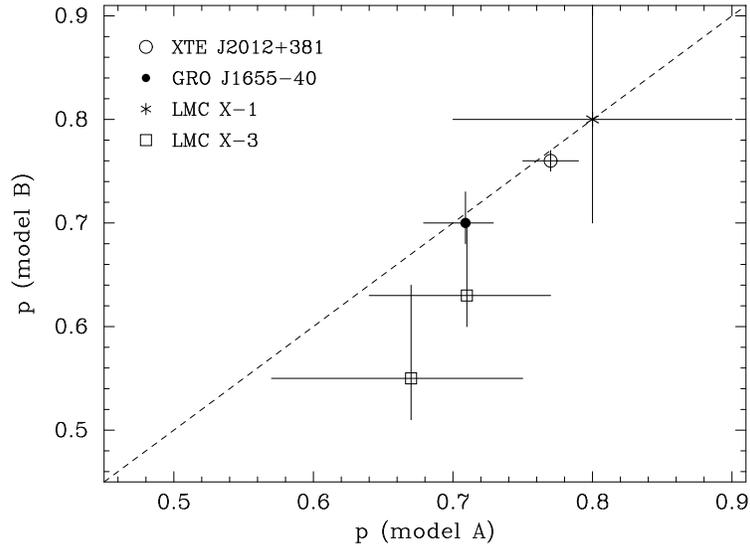}
\caption{Best-fit values of $p$ under model $B$ are plotted against those under 
model $A$.}
\label{fig:p1-p2}
\end{center}
\end{figure}

\begin{figure}[hbtp]
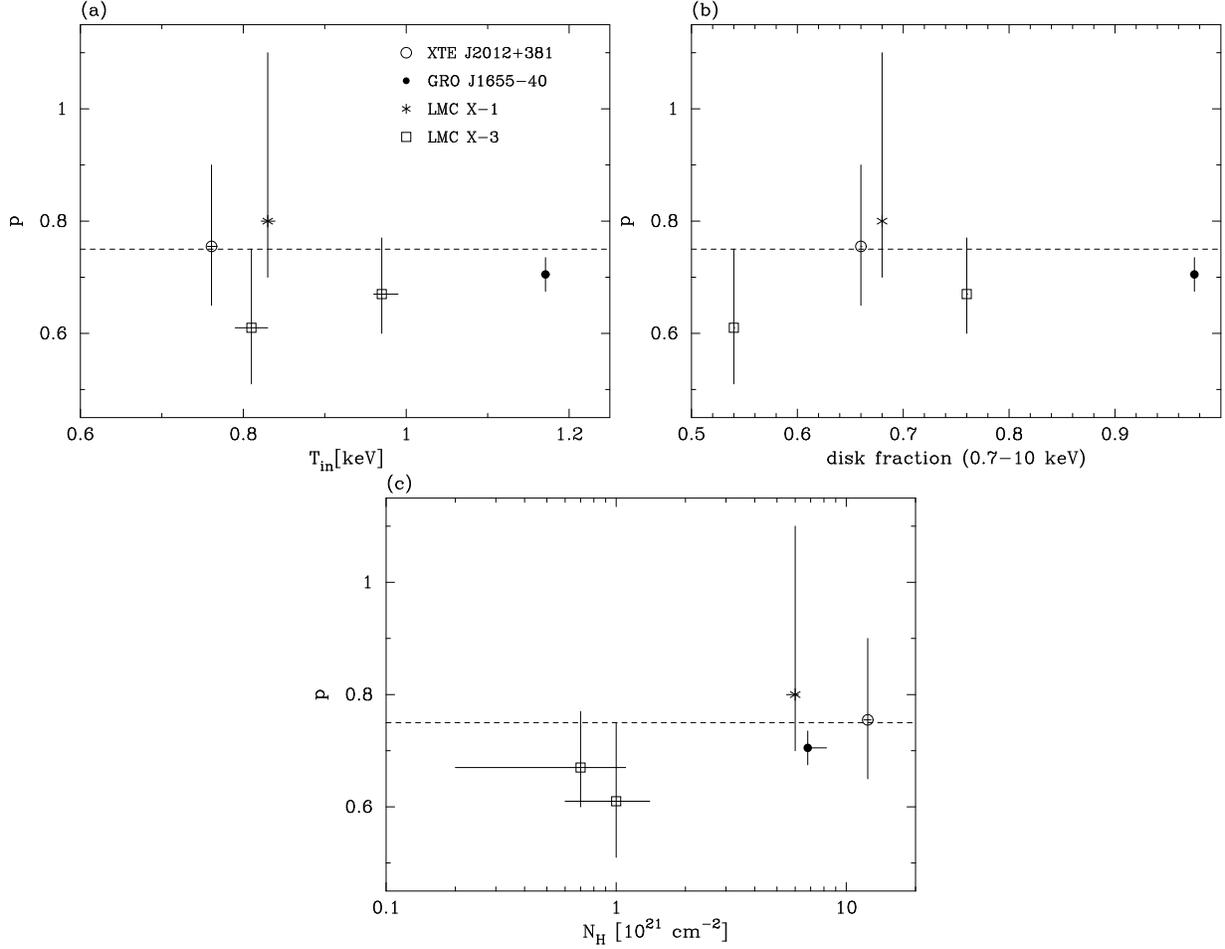

\begin{center}
\includegraphics[scale=.33, angle=-90]{f6a.eps}
\includegraphics[scale=.33, angle=-90]{f6b.eps}
\includegraphics[scale=.33, angle=-90]{f6c.eps}
\caption{Result of fitting with the $p$-free disk model. The best-fit $p$-values 
were plotted against $kT_{\rm in}$ (a), the 0.7--10~keV disk fraction(b), and
 $N_{\rm H}$(c).
In panel $(a)$, $T_{\rm in}$ under the MCD fit is applied instead under the $p$-free disk fit, 
to escape any systematic coupling between the parameters of the $p$-free disk model.
The average of the best fit values of $p$ with model $A$ and model $B$ 
are plotted with marks. 
Each horizontal axis is based on the MCD fit.  
}
\label{fig:p1}
\end{center}
\end{figure}

The disk temperature profiles of four black hole binaries
(XTE~J$2012+381$, GRO~J$1655-40$, LMC~X-1 and LMC~X-3) 
were examined with the $p$-free disk model. 
The uncertainties of the contribution of the hard power-law component were considered by 
fitting the data with two cases, model~$A$ and $B$. 
In Fig.~\ref{fig:p1-p2}, the best-fit values of $p$ under model $B$ are plotted against 
those under model $A$ for our sample.
The obtained values of $p$ are almost the same between these two cases. 
Hereafter, an average of the best-fit values of $p$ with model $A$ and $B$ is 
applied as the most reliable value with its error, which covers an acceptable range of $p$ with 
these two models.
In Fig.~\ref{fig:p1}$a$, the resultant $p$-values are plotted against  $T_{\rm in}$.
Here, $T_{\rm in}$ under the MCD fit is applied instead under the $p$-free disk fit, 
to escape from any systematic coupling between the parameters of the $p$-free disk model.
For MCD temperatures of $T_{\rm in}$=0.76--1.17~keV, 
the central values of $p$ are found to be 
between 0.6 and 0.8 with an average of $0.70\pm0.05$;
the standard $p$-value of $3/4$ is accepted for all sources.

Figure~\ref{fig:p1}$b$-$c$ show other scatter plots, where the obtained values of $p$ are 
plotted against  the 0.7--10~keV disk fraction and $N_{\rm H}$.
The same as in Fig.~\ref{fig:p1}$a$, 
the values of the disk fraction and $N_{\rm H}$ are 
based on the canonical MCD plus power-law model.
In this figure, there is no dependence of $p$ on the other spectral parameters. 
It is thus concluded that 
the observed 0.7--10~keV GIS spectra of the standard high/soft state black hole binaries 
are actually very consistent with the simple standard accretion disks of $p=3/4$ 
for $T_{\rm in}$=0.76--1.17~keV, the disk fraction of 54--98\%, and 
$N_{\rm H}$=0.7--12$\times10^{21}~{\rm cm^{-2}}$.

\begin{figure}[htbp]
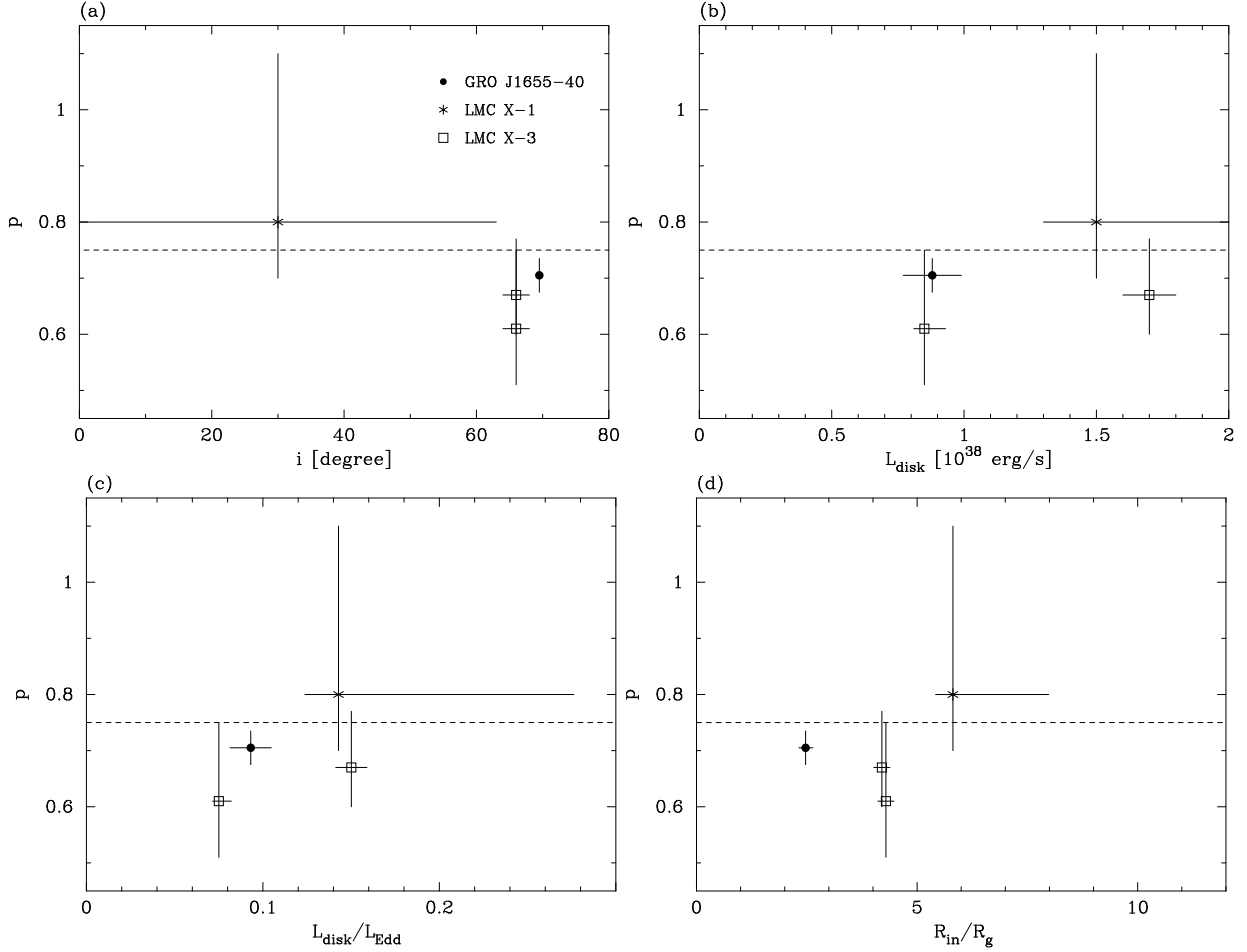

\begin{center}
\includegraphics[scale=.33, angle=-90]{f7a.eps}
\includegraphics[scale=.33, angle=-90]{f7b.eps}
\includegraphics[scale=.33, angle=-90]{f7c.eps}
\includegraphics[scale=.33, angle=-90]{f7d.eps}
\caption{Same as Fig.~\ref{fig:p1}, but the $p$-values are plotted against 
$i$ ($a$), $L_{\rm disk}$ ($b$), $L_{\rm disk}/L_{\rm Edd}$ ($c$), and $R_{\rm in}/R_{\rm g}$ ($d$). 
The errors of the horizontal axes include the systematic uncertainties of $i$ and $d$ together with 
the statistic errors.
}
\label{fig:p2}
\end{center}
\end{figure}

To check the $p$-dependence on the disk luminosity, Eddington ratio, and black hole spin, 
in Fig.~\ref{fig:p2} the values of $p$ are plotted against 
$i$, $L_{\rm disk}$, $L_{\rm disk}/L_{\rm Edd}$, and $R_{\rm in}/R_{\rm g}$.
The result of XTE~J$2012+381$ is not shown in this figure because of a lack of optical information.
Together with 90\% statistical errors, the systematic uncertainties of $i$ and $d$ are considered to
estimate the errors of $L_{\rm disk}$, $L_{\rm disk}/L_{\rm Edd}$ and $R_{\rm in}/R_{\rm g}$.
This figure also does not show any dependency of the $p$-values on the other parameters.

\placefigure{fig:p1-p2}
\placefigure{fig:p1}
\placefigure{fig:p2}

\subsection{On the inner boundary condition and the general relativistic effects}

The standard accretion disk is the basic model for a radiatively efficient, geometrically thin disk. 
A $p$-value of $3/4$ is approximately expected for the standard accretion disk.
However, 
the combination of the boundary condition of the inner edge of the disk, 
the general relativistic effects, 
and the disk vertical structures was expected to change the $p$-value from $3/4$
near the inner edge. 
Therefore, it is meaningful to compare the observed $p$-values 
with those predicted theoretically for the standard disk. 

%
As described in \S~1,  the disk inner-boundary condition of friction free makes 
the temperature profile flatter than $r^{-3/4}$ 
(i.e., the value of $p$ becomes smaller than $3/4$; see Fig.~\ref{fig:tr}). 
If the Schwarzschild potential is applied instead of the Newtonian potential, 
the difference of the temperature profile around the inner boundary becomes 
more significant. 
A general relativistic accretion disk model ({\sc grad} model in the {\sc xspec}; 
Ebisawa, Mitsuda \& Hanawa~1993; Hanawa~1989)
calculates the local effective temperature of the standard accretion disk in the Schwarzschild potential as
{\small 
\begin{equation}
T_{\rm eff}(r)=\left[\frac{3GM\dot{M}}{8\pi\sigma r^3}\cdot\left(1-{\frac{3R_{\rm g}}{r}}\right)^{-1}\cdot \left\{1-\sqrt{\frac{6R_{\rm g}}r}+\sqrt{\frac{3R_{\rm g}}{4r}}\cdot  \ln \left(\frac{1+\sqrt{3R_{\rm g}/r}}{1-\sqrt{3R_{\rm g}/r}}\cdot \frac{1-1/\sqrt{2}}{1+1/\sqrt{2}}\right) \right\} \right]^{1/4}~~~.
\label{eq:tr_s}
\end{equation}
}
In Fig.~\ref{fig:tr}, this temperature profile is shown 
together with that in the Newtonian metric via Eq.~(\ref{eq:tr_n}).
The figure shows that the flattening of the temperature profile at the inner boundary 
is emphasized under the Schwarzschild potential.

In addition to the difference of $T_{\rm eff}(r)$, the gravitational redshift and the 
Doppler boosting are the main relativistic effect; the latter appears most clearly in the dependence of 
the observed spectra on the inclination angle, $i$. 
In order to test both the inner boundary condition and the general relativistic effects in 
the Schwarzschild metric, 
the $p$-values are calculated 
based on the {\sc grad} spectra under the 0.7--10~keV GIS response matrices.
In detail, many {\sc grad} spectra are simulated for several conditions of $i$ and $\dot{M}$, 
and are fitted with the $p$-free disk model and the MCD model. 
To compare the observational results shown in Fig.~\ref{fig:p1}$a$, 
we plot the predicted $p$-values for the 
{\sc grad} model against $T_{\rm in}$ for several conditions of $i$ in Fig.~\ref{fig:alpha:i-t}$a$-$b$.
%
The uncertainty of $N_{\rm H}$ is ignored and considered in panel $(a)$ and $(b)$, respectively.
The trends of $p$-dependence on $T_{\rm in}$ are the same between these two panels, 
though the absolute $p$-values in panel $(b)$ are distributed more widely than those in panel $(a)$. 
In the case of  face-on systems with $i=0^\circ \sim 30^\circ$, 
panel $(b)$ shows that the values of $p$ are expected to appear in the range of 0.68--0.75, 
and do not depend on $T_{\rm in}$ very much.
On the contrary, those for edge-on systems with $i=30^\circ \sim70^\circ$ 
depend on $T_{\rm in}$ significantly, because the gravitational boosting at the inner boundary 
becomes efficient.  
In this case, the values of $p$ are distributed between 0.58--0.7. 

These results show that 
the relativistic effects under the Schwarzschild metric and the boundary condition 
make $p$ {\it smaller} than $3/4$ up to $\Delta p \approx -0.2$.
By comparing Fig.~\ref{fig:p1}$a$ and Fig.~\ref{fig:alpha:i-t}$b$, 
in Fig.~\ref{fig:pg-p}, the observed values of $p$ are plotted against 
those predicted for each binary system based on the {\sc grad} model.
Here, the {\sc grad} predictions were obtained based on the observed 
MCD temperature and the optical information of $i$ for each source by using Fig.~\ref{fig:alpha:i-t}$b$.
For XTE~J$2012+381$ and LMC~X-1 without (or with poor) information on $i$, 
the errors of $p$-values for the {\sc grad} prediction cover the range for $i=0-70^\circ$ and 
$i=0-63^\circ$, respectively, 
and their central values are just taken from the center for the confidential range. 

That the observed best-fit values of $p$ are consistent with those calculated 
for the {\sc grad} model, and
a trend toward a slightly larger value of the observed-$p$ than the {\sc grad}-$p$  
($\Delta p\approx +0.06$) is marginally suggested.

\placefigure{fig:alpha:i-t}
\placefigure{fig:pg-p}

\subsection{On the disk vertical structure}

\begin{figure}[hbtp]
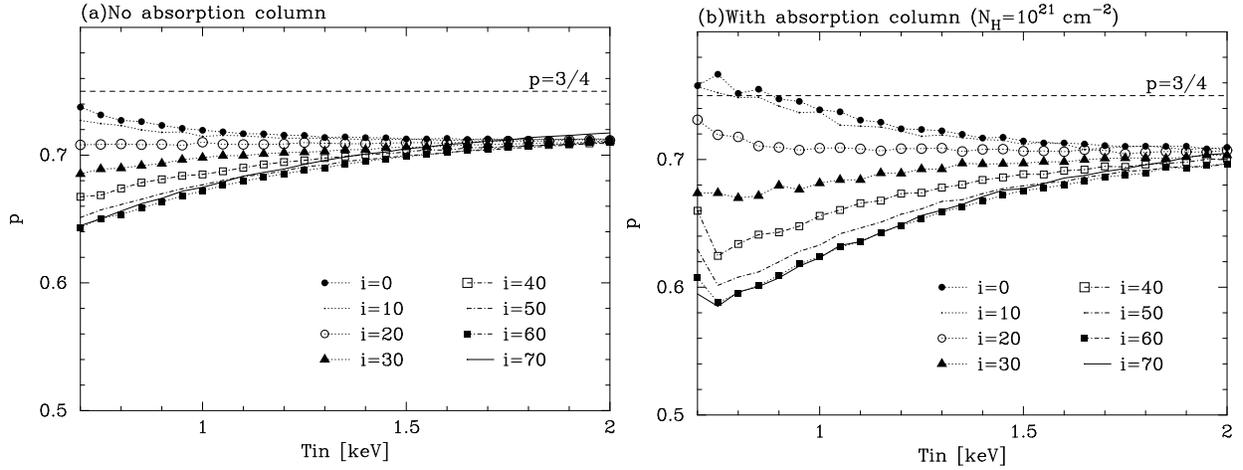

\begin{center}
\includegraphics[angle=-90,scale=.33]{f8a.eps}
\includegraphics[angle=-90,scale=.33]{f8b.eps}
\caption{Expected values of $p$ against $T_{\rm in}$ 
for the {\sc grad} spectra with several inclination angles, $i$.
The horizontal axis, $T_{\rm in}$, is based on the MCD fit. 
The uncertainty of the systematic coupling of the absorption column
and the $p$-value is ignored in panel $(a)$ and considered in panel $(b)$. 
The detailed criteria for the simulation are as follows. 
Panel(a)---(1)Many {\sc grad} spectra without $N_{\rm H}$ 
are calcurated by changing both $i$ and $\dot{M}$.
The simulated spectra are convoluted with the response function of the
0.7--10~keV GIS by including 1\% systematic errors.
(2) The faked {\sc grad} spectra were fitted with both the MCD model and 
the $p$-free disk model without $N_{\rm H}$. 
(3) For  the same {\sc grad} spectra, the best-fit $p$ of the $p$-free disk model 
are plotted against the best-fit $T_{\rm in}$ of the MCD model. 
Panel (b)---Same as panel $(a)$, but the spectra were simulated and fitted by 
including $N_{\rm H}$. 
The plot is the result for $N_{\rm H}$ of $\approx 10^{21}~{\rm cm^{-2}}$. 
This plot is not influenced by changing $N_{\rm H}$ in the range of $10^{20-22}~{\rm cm^{-2}}$. 
}
\label{fig:alpha:i-t}
\end{center}
\end{figure}

\begin{figure}
\begin{center}
\includegraphics[angle=-90,scale=.4]{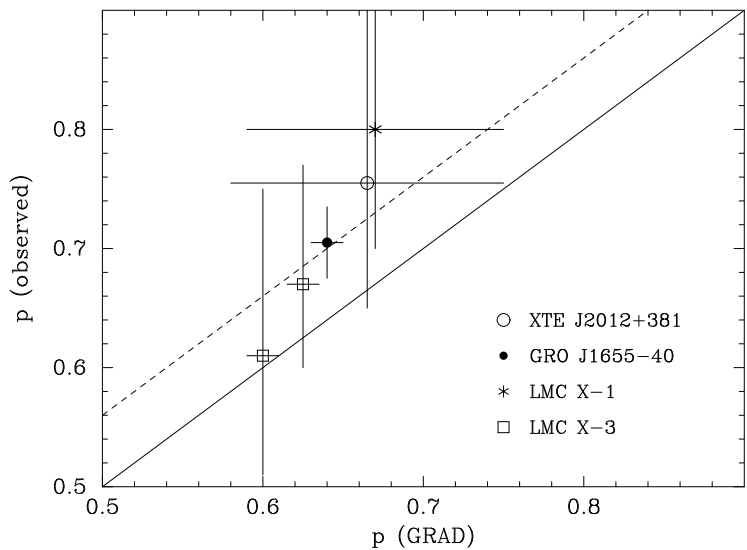}
\caption{Obtained values of $p$ plotted against 
the expected values of $p$ for each binary system 
based on the {\sc grad} model in Fig.~\ref{fig:alpha:i-t} with information of $i$ and $T_{\rm in}$. 
The errors of $p$ of
 the {\sc grad} expectation include the uncertainty of the estimation of $i$. 
 Especially for XTE~J$2012+381$, the range of $p$ under the {\sc grad} prediction is justified 
 for $i$ of 0--70$^\circ$.
The solid line and dashed line mean
 $p({\rm observed})=p$({\sc grad}) and $p({\rm observed})=p$({\sc grad})$+0.06$, 
 respectively.}
\label{fig:pg-p}
\end{center}
\end{figure}

The consistency of the $p$-values between the observations and the {\sc grad} calculations  
means that the disk vertical structure, which is not included in the {\sc grad} model, 
should not change the spectral shape very much. 
Therefore, the assumption concerning the sum of the local diluted blackbody 
radiations with a radially constant hardening factor, $\kappa$, is justified for 
black hole binaries in the standard high/soft state.
Otherwise, the vertical structure can shift the $p$-value to slightly {\it larger}, as 
marginally suggested in Fig.~\ref{fig:alpha:i-t}.

Is this observational result consistent with the current models for a 
geometrically thin disk that includes the disk vertical structure?
Even within the standard disk picture, electron scattering in the disk 
direction is expected to be important.
Since electron scattering under optically thick conditions usually becomes more 
important for higher temperature (e.g., Rybicki \& Lightman 1979), 
emission from the higher temperature inner part of the disk may be suppressed more than 
the lower temperature outer part. Thus, with electron scattering, 
the $p$-value roughly becomes smaller than the standard value of $3/4$.
The real disk is much more complicated, and thus such a 
brief estimation may not be realistic.
In fact, the disk spectra that take account for electron scattering 
have been computed by many authors, including
Shimura \& Takahara 1995, Ross \& Fabbian (1996), and Merloni, Fabbian \& Ross (2000), 
on many conditions.


Recently, significant theoretical advances have been made in understanding the nature and the
vertical structure of the optically-thick and geometrically-thin disk. 
Turner~(2004) shows that magnetic stress should play an important role in the disk, 
and is considered to make the radiatively efficient disk stable. 
Bound-free metal opacity is also thought to be important (Davis et al.~2005).
Together with these theoretical studies, the observationally determined temperature profile 
for the disk can be  a good
probe to see the distortion of the spectral shape due to the disk vertical structure.


\placefigure{fig:alpha:st95}

\acknowledgments

We are grateful to T. Shimura for comments on the disk model by Shimura \& Takahara (1995), and
A. K would like to thank C. Done, S. Mineshige and Y. Terashima for valuable discussions.
A.K. is supported by a special postdoctoral researchers program in RIKEN.

\clearpage
\begin{deluxetable}{cccccrr}
\tabletypesize{\scriptsize}
\tablecaption{Observations with {\it ASCA} and optical data of our sample.
\label{tab:obs}}
\tablewidth{0pt}
\tablehead{
\colhead{object	}& \colhead{BH mass}  &\colhead{inclination} & \colhead{Distance}& \colhead{observation date} & \colhead{exp.}& \colhead{ref$^a$} \\
\colhead{}	&\colhead{$M_\odot$} &\colhead{[degree]}& \colhead{[kpc]}&\colhead{yyyy/mm/dd}&\colhead{[ks]}&\colhead{}
}

\startdata
XTE~J$2012+381$& ---	&---	&--- &1998/5/29-30 &2.8 &1,2,3\\
GRO J$1655-40$ &	6.0--6.6	&$69.5\pm0.08$&$3.2\pm0.2$	&1997/2/26	&2.4	&4,5,6,7,8,9,10\\
LMC X-1	&4--10	&$<63$	&55 &1995/4/2-3	&21.7	&11,12,13,14		\\
LMC X-3	&5.9--9.2	&64--68	&55 &1993/9/22-23	&20.6	&11,14,15		\\
	&	&	& 	&1995/4/14-15	&19.9	&		\\
\enddata
\tablenotetext{a}{{\sc References}--- 
$^1$White et al.~1998; $^2$Vasiliev et al.~2000; $^3$Hynes et al.~1999;
$^4$Hjellmiung \& Ruppen (1995); $^5$Orosz \& Bailyn (1997); $^6$Shahbaz et al.(1999); 
$^7$Zhang et al. (1997); $^8$Ueda et al.~(1998); $^9$Yamaoka et al.~(2001); $^{10}$Kubota et al. (2001)
$^{11}$Feedman et al.~(2001);  $^{12}$Hutchings et al.~(1987); $^{13}$Cowley et al.~(1995); $^{14}$Makishima et al.~(2000);
$^{15}$Cowley et al.~(1983); 
}

\end{deluxetable}

\begin{table}[hbpt]
\begin{center}
\caption{Best-fit parameters obtained by a simultaneous {\it ASCA}/{\it RXTE} observation of 
XTE J$2012+381$.}
\label{tab:result_j2012}
{\footnotesize
\begin{tabular}{ccccccc}
\hline \hline
\multicolumn{2}{l}{parameters}	&MCD	&\multicolumn{2}{c}{$p$-free disk}\\
	&	&	&$A$	&$B$	\\
\hline
\multicolumn{2}{l}{$N_{\rm H}$ $(10^{22}~{\rm cm^{-2}})$} & $1.24 ^{+0.04}_{-0.05}$	& $1.19 \pm 0.05$ &$1.1\pm0.1$	\\
\multicolumn{2}{l}{MCD or $p$-free disk}	&	&	&\\
	&$kT_{\rm in}$ (keV)	&$0.761\pm0.005$&$0.77\pm0.02$	&$0.76\pm0.01$	\\
	&norm. ($\times 10^3$)$^a$	&$1.62\pm0.05$	&$1.4^{+0.4}_{-0.3}$&	$2.0^{+0.8}_{-0.5}$\\
	&$p$			&(0.75)	&$0.71\pm0.06$	&$0.8\pm0.1$	\\
\multicolumn{2}{l}{power-law}	&	&	&	\\
	&$N_{\rm H}^{\rm hard}$ $(10^{22}~{\rm cm^{-2}})$&---	&---	&$25\pm8$\\
	&$\Gamma$		&$2.36^{+0.06}_{-0.07}$&$2.3\pm0.1$&$2.6\pm0.1$\\
	&norm.$^b$		&$0.44\pm0.07$	&$0.4\pm0.1$&$1.0^{+0.4}_{-0.3}$\\
\multicolumn{2}{l}{Absorption edge}&	&		&\\
	&$E$ (keV)		&$7.6^{+0.1}_{-0.2}$	&$7.3\pm0.3$	&$8.3^{+0.2}_{-0.3}$	\\
	&$\tau$		&$>0.9$	&$2^{+4}_{-1}$&$>0.8$\\
	&width (keV)		&$2.7^{+0.4}_{-0.5}$	&$5^{+19}_{-2}$&$2.6^{+0.4}_{-0.7}$	\\
\hline
\multicolumn{2}{l}{$\chi^2/\nu$}&152.1/152	&151.1/151	&139.7/150\\
\hline
\multicolumn{2}{l}{disk fraction$^c$}	&0.66	& 0.67	&0.95\\
\multicolumn{2}{l}{$R_{\rm in}$ [km]$^{de}$}	&$68\pm1$	& ---	&---\\
\multicolumn{2}{l}{$L_{\rm disk}~[10^{38}~{\rm erg~s^{-1}}]^d$}	&1.4	& ---	&---\\
\hline
\multicolumn{5}{l}{{\sc Note}--- Errors represent 90 \% confidence limits. }\\
\multicolumn{5}{l}{$^a$ In unit of $(r_{\rm in}/{\rm km})^2\cdot \cos i\cdot (d/10~{\rm kpc})^{-2}$}\\
\multicolumn{5}{l}{$^b$ In unit of ${\rm photons~s^{-1}~cm^{-2}~keV^{-1}}$ at 1 keV}\\
\multicolumn{5}{l}{$^c$ A ratio of flux of the MCD component to the total flux in 0.7--10~keV.}\\
\multicolumn{5}{l}{$^d$ Estimated for assumed distance of 10~kpc and inclination of $60^\circ$. }\\
\multicolumn{5}{l}{$^e$ With $\kappa=1.7$ and $\xi=0.41$}\\
\end{tabular}
}
\end{center}
\end{table}

\begin{table}[hbpt]
\begin{center}
\caption{Same as Table~\ref{tab:result_j2012}, but for GRO J1655-40.}
\label{tab:result_j1655}
{\footnotesize
\begin{tabular}{clcccc}
\hline \hline
\multicolumn{3}{l}{parameters}&MCD&\multicolumn{2}{c}{$p$-free disk}\\
	&	&	&   &$A$ &$B$	\\
\hline
\hline
\multicolumn{3}{l}{$N_{\rm H}$ $(10^{21}~{\rm cm^{-2}})$} &$6.8^{+1.4}_{-0.2}$ &$7.1^{+0.2}_{-0.3}$& $7.1\pm0.3$\\
\multicolumn{3}{l}{disk}	&			&&\\
	&\multicolumn{2}{c}{$kT_{\rm in}$ (keV)}&$1.171^{+0.005}_{-0.006}$&$1.19\pm0.01$&$1.20^{+0.01}_{-0.02}$\\
	&\multicolumn{2}{c}{norm. ($\times 10^3$) }&$1.21^{+0.02}_{-0.04}$	&$1.0\pm0.1$&$0.97^{+0.07}_{-0.13}$	\\
	&\multicolumn{2}{c}{$p$}	&		(0.75)	&$0.71^{+0.02}_{-0.03}$&$0.70^{+0.03}_{-0.02}$\\
\multicolumn{3}{l}{power-law}	&&		&\\
	&\multicolumn{2}{c}{$N_{\rm H}^{\rm hard}$ $(10^{24}~{\rm cm^{-2}})$}&---	&---& $7^{+8}_{-7}$\\
	&\multicolumn{2}{c}{$\Gamma$}	&$2.7^{+0.3}_{-0.5}$	&$(2.2^{>+0.3}_{<-0.2})$	&$(2.5^{>+0}_{<-0.5})$\\
	&\multicolumn{2}{c}{norm}		&$0.5^{+3.2}_{-0.4}$&$0.13^{+0.18}_{-0.06}$ &$0.35^{+0.07}_{-0.28}$	\\
\multicolumn{3}{l}{Absorption structures$^a$}&	&	&\\
	&edge1$^b$ &$\tau$		&$0.15\pm0.03$	&$0.19\pm0.04$&$0.18^{+0.04}_{-0.02}$	\\
	&edge2$^b$ &$\tau$		&$0.21^{+0.04}_{-0.05}$	&$0.22^{+0.04}_{-0.05}$&$0.21^{+0.07}_{-0.05}$	\\
	&edge3 &$E$~[keV]&$10.4\pm0.2$	&$10.4^{+0.1}_{-0.2}$&$10.4^{+0.1}_{-0.3}$	\\
	&&$\tau$		&$0.27\pm0.05$	&$0.29^{+0.04}_{-0.07}$&$0.28^{+0.05}_{-0.08}$	\\
	&gaussian$^c$ &EW	 [eV]	&$61\pm18$	&$73^{+16}_{-23}$&$79\pm18$	\\
\hline
\multicolumn{3}{l}{$\chi^2/\nu$}&123.1/150&116.4/149	&114.8/148	\\
\hline \hline
\multicolumn{3}{l}{disk fraction}&0.975	&0.990	&0.999\\
\multicolumn{3}{l}{$R_{\rm in}$ [km] $^{de}$}	&$22.6\pm0.2~^{+1.5}_{-1.4}$	& ---	&---\\
\multicolumn{3}{l}{$L_{\rm disk}~[10^{38}~\rm{erg~s^{-1}}]$ $^{df}$}	&$0.88\pm0.11$	& ---	&---\\
\hline
\multicolumn{6}{l}{$^a$ Referring to Yamaoka et al.~(2001). }\\
\multicolumn{6}{l}{$^b$ Energies of  edge1 and edge2 are fixed at 7.71~keV and 8.81~keV, }\\
\multicolumn{6}{l}{~~respectively.}\\
\multicolumn{6}{l}{$^c$ Negative gaussian to represent an absoroption line.  Line center}\\
\multicolumn{6}{l}{~~energy and $\sigma$ are fixed at 
6.81~keV and 160~eV, respectively. }\\ 
\multicolumn{6}{l}{$^d$ Central values are estimated for $d=3.2$~kpc and $i=70^\circ$. }\\
\multicolumn{6}{l}{$^e$ Errors include the statistic error (first one) and the uncertainties of }\\
\multicolumn{6}{l}{~~~$i$ and $d$ (second one).}\\
\multicolumn{6}{l}{$^f$ Errors include the uncertainties of $i$ and $d$.}\\
\end{tabular}
}
\end{center}
\end{table}

\begin{table}[hbpt]
\begin{center}
\caption{Same as Table~\ref{tab:result_j1655}, but for an {\it ASCA} observation of LMC X-1.}
\label{tab:result_lmcx1}
{\footnotesize
\begin{tabular}{ccccc}
\hline \hline
\multicolumn{2}{l}{parameters}	&MCD	&\multicolumn{2}{c}{$p$-free disk}\\
	&	&	&$A$	&$B$	\\
\hline
\multicolumn{2}{l}{$N_{\rm H}$ $(10^{21}~{\rm cm^{-2}})$} &$6.0^{+2}_{-5}$ 	& $6.0^{+0.2}_{-0.5}$ &$ 4.4^{+1.6}_{-0.4} $	\\
\multicolumn{2}{l}{disk}	&	&	&\\
	&$kT_{\rm in}$ (keV)&$0.830^{+0.009}_{-0.008}$&$0.81^ {+0.07}_{-0.02}$	&$0.82\pm0.04$\\
	&norm&$85^{+6}_{-5}$	&$ 103^{+34}_{-44}$&	$112^{+78}_{-41}$\\
	&$p$			&(0.75)	&$ 0.8\pm0.1$	&$0.8^{+0.3}_{-0.1}$	\\
\multicolumn{2}{l}{power-law}	&	&	&\\
	&$N_{\rm H}^{\rm hard}$ $(10^{22}~{\rm cm^{-2}})$&---	&---	&$0.5^{+2.5}_{-0.5}$\\	
	&$\Gamma$		&$2.4\pm0.2$	&($2.5^{>+0}_{<-0.5}$)&($2.5^{>+0}_{<-0.5}$)\\
	&norm		&$0.11^{+0.04}_{-0.02}$	&$0.14\pm^{+0.1}_{-0.07}$&$0.137^{+007}_{-0.06}$\\
\multicolumn{2}{l}{Absorption edge}&	&	&\\
	&$E$ (keV)		&$7.2^{+0.4}_{-0.3}$	&$7.1\pm0.4$	&$7.1\pm0.4$	\\
	&$\tau$		&$>0.6$&$>0.5$	&$>0.5$	\\
	&width (keV)		&$2\pm1$	&$2\pm1$&$2\pm1$	\\
\hline
\multicolumn{2}{l}{$\chi^2/\nu$}&94.6/110	&93.7/109&92.8/108  \\
\hline \hline
\multicolumn{2}{l}{disk fraction}&0.68	&0.64	&0.69\\
\multicolumn{2}{l}{$R_{\rm in}$ [km]$^a$}	&$59\pm2~^{+22}_{-4}$	& ---	&---\\
\multicolumn{2}{l}{$L_{\rm disk}~[10^{38}~\rm{erg~s^{-1}}] ^a$}	&$1.5^{+1.4}_{-0.2}$	& ---	&---\\ \hline
\multicolumn{5}{l}{$^a$ Central values are estimated for $d=50$~kpc and $i=30^\circ$. }\\
\end{tabular}
}
\end{center}
\end{table}

\begin{table}[hbpt]
\begin{center}
\caption{Same as Table~\ref{tab:result_j1655}, but for {\it ASCA} observations of LMC X-3.}
\label{tab:result_lmcx3}
{\footnotesize
\begin{tabular}{cc|ccc|ccc}
\hline \hline
&&\multicolumn{3}{c|}{1993}&\multicolumn{3}{c}{1995}\\
\multicolumn{2}{l|}{parameters}	&MCD	&\multicolumn{2}{c|}{$p$-free disk}	&MCD&\multicolumn{2}{c}{$p$-free disk}\\
	&	&	&$A$	&$B$ & &$A$ &$B$	\\
\hline
\hline
\multicolumn{2}{l|}{$N_{\rm H}$ $(10^{21}~{\rm cm^{-2}})$} 
&$1.0\pm0.4$ 	& $ 0.8^{+0.1}_{-0.5}$ &$0.9^{+0.1}_{-0.9}$
&$0.7^{+0.4}_{-0.5}$ & $0.4^{+0.1}_{-0.4}$&	$3\pm2$\\
\multicolumn{2}{l|}{disk}	&	&	&\\
&$kT_{\rm in}$ (keV)&$0.81\pm0.2$&$0.83^{+0.07}_{-0.03}$	&$0.83\pm0.05$
&$0.97^{+0.2}_{-0.1}$ & $0.98^{+0.04}_{-0.02}$ & $0.99\pm0.04$\\
&norm&$25\pm3$	&$19\pm8$&	$14^{+15}_{-5}$
&$24\pm2$ & $21\pm6$&$16^{+6}_{-3}$\\
	&$p$			&(0.75)	&$0.67^{+0.08}_{-0.10}$	&$0.55^{+0.09}_{-0.04}$	&(0.75)&$0.71^{+0.06}_{-0.07}$&$0.63^{+0.07}_{-0.03}$\\
\multicolumn{2}{l|}{power-law}	&	&	&&&&\\
	&$N_{\rm H}^{\rm hard}$ $(10^{22}~{\rm cm^{-2}})$&---	&---	&$3^{+3}_{-1}$&---&---&$3^{+3}_{-1}$\\	
	&$\Gamma$		&$2.7^{+0.1}_{-0.1}$	&($2.5^{>+0}_{<-0.5}$)&($2.5^{>+0}_{-0.4}$)&$2.7^{+0.2}_{-0.3}$&($2.5^{>+0}_{<-0.5}$)&($2.5^{>+0}_{<-0.5}$)\\
	&norm. [$\times10^{-2}$]	&$6 \pm 2$	&$4^{+1}_{-3}$ & $5^{+4}_{-3}$ 
	& $5\pm2$ & $3^{+1}_{-2}$ & $3^{+6}_{-2}$ \\
\hline
\multicolumn{2}{l|}{$\chi^2/\nu$}&102.2/109	&102.6/108&98.9/107
&103.8/109&104.0/108&101.4/107\\
\hline \hline 
\multicolumn{2}{l|}{disk fraction}	&0.54	&0.63	&0.86&0.76&0.82&0.94\\	
\multicolumn{2}{l|}{$R_{\rm in}$~[km]$^a$}	&$47\pm2~\pm2$	& ---	&---&$46\pm2~\pm2$	& ---	&---\\
\multicolumn{2}{l|}{$L_{\rm disk}~[10^{38}~\rm{erg~s^{-1}}]^a$}	&$0.85^{+0.08}_{-0.04}$	& ---	&---&$1.7\pm0.1$ & --- & ---\\
\hline
\multicolumn{8}{l}{$^a$ Central values are estimated for $d=50$~kpc and $i=66^\circ$. }\\
\end{tabular}
}
\end{center}
\end{table}

\clearpage

\end{document}